\documentclass[twocolumn]{aastex631}

\newcommand\kms{\rm km\ s^{-1}}

\usepackage{float}
\usepackage{parskip}

\shorttitle{Runaway OB Stars in the SMC}
\shortauthors{Phillips et al.}

\graphicspath{{./}{figures/}}

\begin{document}

\title{Runaway OB Stars in the Small Magellanic Cloud III. Updated Kinematics and Insights on Dynamical vs Supernova Ejections}

\correspondingauthor{M. S. Oey}
\email{msoey@umich.edu}

\author[0000-0002-8086-5906]{Grant D. Phillips}
\affiliation{Astronomy Department, University of Michigan, 1085 South University Avenue, Ann Arbor, MI 48109-1107, USA }

\author[0000-0002-5808-1320]{M. S. Oey}
\affiliation{Astronomy Department, University of Michigan, 1085 South University Avenue, Ann Arbor, MI 48109-1107, USA }

\author{Maria Cuevas}
\affiliation{Private address}

\author{Norberto Castro}
\affiliation{Leibniz-Institut f\"ur Astrophysik Potsdam (AIP), An der Sternwarte 16, 14482, Potsdam, Germany}
\altaffiliation{
Present address:  Institut f\"ur Astrophysik, Georg-August-Universit\"at, Friedrich-Hund-Platz
1, 37077 G\"ottingen, Germany}

\author{Rishi Kothari}
\altaffiliation{Present address:  Private}
\affiliation{Astronomy Department, University of Michigan, 1085 South University Avenue, Ann Arbor, MI 48109-1107, USA }

\submitjournal{The Astrophysical Journal}
\received{January 10, 2024}
\accepted{March 24, 2024}

\begin{abstract}

We use the kinematics of field OB stars to estimate the frequencies of runaway stars generated by the dynamical ejection scenario (DES), the binary supernova scenario (BSS), and the combined two-step mechanism.
We update the proper motions for field OB and OBe stars in the Small Magellanic Cloud (SMC) using Gaia DR3.  Our sample now contains 336 stars from the Runaways and Isolated O-Type Star Spectroscopic Survey of the SMC (RIOTS4), and we update our algorithm to calculate more accurate velocities compared to those obtained previously from DR2. We find a decrease in median velocity from 39 to 29 $\kms$, implying that the proper motions from our previous work were systematically overestimated. We present the velocity distribution for OBe stars and quantitatively compare it to those of non-compact binaries and high-mass X-ray binaries. We confirm that OBe stars appear to be dominated by the BSS and are likely post-SN binary systems, further supporting the mass-transfer model to explain the origin of their emission-line disks. In contrast, normal OB stars may show a bimodal velocity distribution,
as may be expected from different processes that occur with dynamical ejections.
The kinematics of fast-rotating OB stars are similar to those of normal OB stars rather than OBe stars, suggesting that the origin of their high $v_r\sin i$ is different from that of OBe stars.  
We update our model parameters describing the kinematic origins of the SMC field population, still confirming that for runaway stars, the DES mechanism dominates, and two-step ejections seem comparable in frequency to pure BSS ejections.
\end{abstract}


\setlength\parindent{5pt}
\setlength{\parskip}{2mm}

\section{Introduction} \label{sec:Introduction}

The kinematics of massive field stars have recently confirmed that almost all of them are ejected from their parent clusters \citep[e.g.,][]{PaperI,DorigoJones2020}, with $\lesssim 5$\% likely to have formed as relatively isolated field OB stars \citep[e.g.,][]{VargasSalazar2020}. Field stars also constitute a substantial fraction of all massive stars, and in particular, about $\sim20-30$\% of the total OB star population are field stars \cite[]{Oey04}. Thus, the kinematics of field OB stars offer a vital opportunity to probe cluster dynamical evolution, binary population parameters, and stellar evolution.

The two main mechanisms for producing runaway OB stars are \citep{Hoogerwerf01} the dynamical ejection scenario (DES) and the binary supernova scenario (BSS).  These generate distinct velocity distributions \cite[e.g.,][]{Leonard88,Renzo19}. In the DES, a star is ejected from its parent cluster due to a close encounter with a binary or another system. Observations of stellar clusters show mass segregation where the most massive stars tend to be found near the center of a cluster \cite[e.g.,][]{LadaLada03}. Additionally, simulations show that a massive, “bully binary” in the cluster center dominates the cross section for dynamical interactions \cite[]{FujiiPZ}. Together, these findings show that the runaway population for the DES is weighted toward higher-mass stars and that the runaway fraction tends to increase with mass \cite[]{PeretsSubr}. In general, the stars with the largest proper motions result from dynamical ejections.  Binaries can themselves be dynamically ejected as binary runaway systems.

In the BSS, the core-collapse supernova (SN) of the more evolved star in a tight binary system causes the 
acceleration of its companion, which may or may not become unbound from the primary star's compact remnant. 
Both the mass loss from the primary and the explosion kick can give the secondary star a velocity on the order of its original pre-SN orbital velocity \cite[]{Blaauw61,Leonard88}. If the system remains bound, it may be observed as a high-mass X-ray binary (HMXB). Additionally, we believe that OBe stars are predominantly objects that have been spun up by binary mass transfer and then ejected as runaways by the BSS mechanism \citep[e.g.,][]{Dallas2022, DorigoJones2020}; their Balmer emission originates from circumstellar disks that are due to their fast rotation velocities.

A combination mechanism, the two-step ejection, can occur when a system experiences both dynamical and supernova ejection \citep{Pflamm-Altenburg10}.   
If tight binaries are initially dynamically ejected, they are also likely to appear as post-interaction OBe stars after the first SN explosion.
Two-step ejections therefore may be a significant subset of the BSS population \citep{DorigoJones2020}.

In previous work from our group, \citet{PaperI} and \citet{DorigoJones2020} studied the kinematics of OB field stars in the Small Magellanic Cloud (SMC) using proper motion data from Gaia DR2. They found that dynamical ejections dominate over supernova ejections by a factor of $\sim2-3$. They also suggest that the BSS runaway population may be dominated by two-step ejections. The release of DR3 from Gaia \citep{Gaia2023} provides the opportunity to update the findings from these preliminary kinematics. In this work, we recalculate the relative contributions from the DES and BSS in the SMC with our updated set of proper-motion velocities, greatly clarifying the kinematics of massive field stars. We also examine the BSS contribution in more detail, by compiling the velocity distribution of OBe stars and considering fast rotators that are non-OBe stars.

\section{Methods} \label{sec:Methods}

The basis for our stellar sample is The Runaways and Isolated O-Type Star Spectroscopic Survey of the SMC (RIOTS4). This survey by \cite{Lamb16} consists of 374 massive field stars, which are defined to be at least 28 pc away from any other OB candidate. The sample was identified using the $UBVI$ photometry of \citet{Massey02} and spectroscopically confirmed to have stellar spectral types of O and B \citep{Lamb16}.  To calculate new residual velocities, we use the proper motions from \cite{Gaia2023}.  We also use stellar masses and rotational velocities in our analysis that were obtained from RIOTS4 spectra by \citet[][hereafter Paper~I]{PaperI}, \citet[][hereafter Paper~II]{DorigoJones2020}, and \cite{Dallas2022}. 

We remove 16 stars from the sample that have proper motion errors greater than one standard deviation from the median error. We also remove 8 stars that were included in the RIOTS4 sample as supplementary objects, but which did not meet their criteria for field stars \citep{Dallas2022}, as well as 5 stars with no available Gaia data, 4 B[e] stars, 3 stars with spurious Gaia motions, and 2 Wolf-Rayet stars. This brings our final sample down to 336 OB field stars, an increase in size by 11\% compared to Paper I, which had 304 stars. We update some spectral types, mostly for OBe stars, based on new review of the RIOTS4 data; and for stars with poor classifications, we use the compilation by \citet{Dallas2022}.  We have stellar mass data for 334 of our stars: 283 from Paper II, and an additional 51 from Dallas et al. 2022. Additionally, we have $v_{r}\sin(i)$ data for 189 of our stars. In our final sample, there are 142 OBe stars that appear to be single star systems, 16 emission-line HMXBs, 1 non-emission HMXB, 10 double-lined spectroscopic binaries (SB2), 9 eclipsing binaries (EB), and 4 systems that are both EBs and SB2s. 
In what follows, we refer to EBs and SB2s as non-compact binaries.
We note that 2 of our non-compact binaries, [M2002] SMC-24119 and 30744, are emission-line systems, meaning that our total OBe sample contains 160 stars. The remaining 154 stars appear to be single OB stars, although binaries can be difficult to identify. These data and identifications are provided in Table~\ref{tab:Table 1}.

Because non-compact binaries are, by definition, pre-supernova objects, they can only be ejected dynamically. Therefore, we use our sample of EBs and SB2s
to trace objects accelerated dynamically.
Also assuming that our normal OB stars are dynamically ejected (Paper~II), we have
a total of 177 DES objects. Conversely, HMXBs are post-supernova systems since they contain a compact object, either a neutron star or black hole, both of which are remnants of core-collapse supernovae. As argued in Paper~II, the majority of OBe stars are likely also post-supernova binary systems due to their being spun up by binary mass transfer. Therefore, we use our samples of OBe stars and HMXBs to identify 
objects accelerated by supernovae, giving us a total of 159 BSS objects; this excludes the two OBe stars that are non-compact binaries. 
We caution that the above identification of DES and BSS objects is
only valid to first-order; for example, it is likely that our sample of OBe stars is an incomplete representation
of BSS objects, while some OBe stars may not originate from binary mass transfer. We discuss this in the next section.

\begin{figure}[H]
    \includegraphics[scale=0.3] {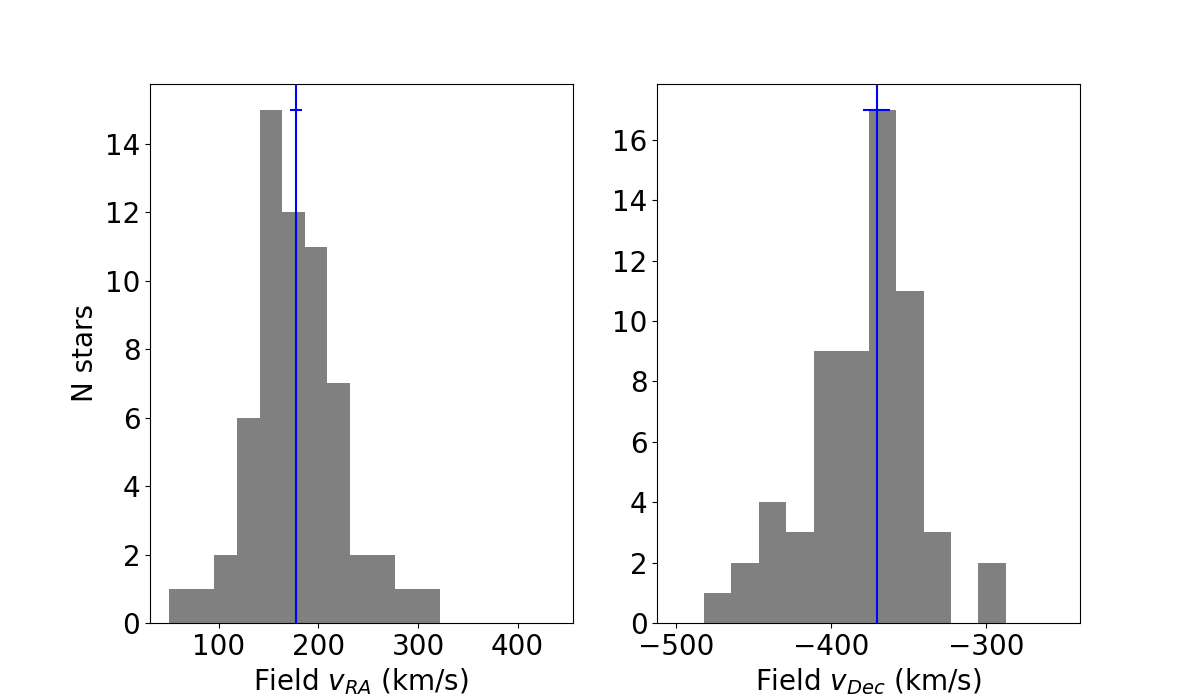}
\caption{RA and Dec proper motion velocity histograms of the stars within $5\arcmin$ of target star [M2002] SMC-1037. Median velocities (blue vertical lines) and errors (blue horizontal lines) of $163.7 \pm 8.1\ \kms$  and $-370.6 \pm 7.3\ \kms$, respectively, are shown. The field contains 55 stars.
\label{fig:Figure 1}}
\end{figure}

The transverse velocities for each target star are calculated relative to its local systemic field motion similarly to Paper~I, but with some refinements as follows. 
We consider the local field stars within a $5\arcmin$ radius of each target star.
In Paper I, the background field star sample only included stars with magnitude $G < 17$ from the catalog of Massey (2002). In this work, we use the Gaia catalog stars, increasing the limit to all detected stars having $G < 18$, allowing
larger field star samples.  We remove outliers whose Gaia proper motions fall outside the limits 50 $\leq$ $v_{\rm RA}$ $\leq$ 450   or $-500\leq v_{\rm Dec}\leq -250\ \kms$ from these local field star samples. We define the local systemic velocities to be the median RA and Dec velocities of these stars within each field, excluding the motion of the target star. Figure \ref{fig:Figure 1} shows an example histogram of the RA and Dec field star velocity distributions for star [M2002]SMC-1037. 

To find the residual RA and Dec velocities of the target star relative to its field systemic motion, we subtract our calculated field velocities from its Gaia proper motion velocities. To calculate the proper motion velocity errors for the target stars, we combine in quadrature its Gaia proper motion measurement errors in RA and Dec with the standard errors of the medians for the corresponding field velocities. We find that our median transverse velocity error decreases by a factor of $\sim1.7$, from $27\ \kms$ in Paper~I to $16\ \kms$ with a standard deviation of $6\ \kms$ in this work, due to both our updated algorithm and improved Gaia measurements. 

The obtained velocities for the sample are presented in Table~\ref{tab:Table 1}.  Columns 1 -- 3 show, respectively, the star ID from \citet{Massey02}, the sub-population, if any, and the spectral type. Columns 4 and 5 give the Gaia DR3 RA and Dec J2016 coordinates, respectively. Columns 6 -- 9 give the target star RA and Dec velocities with their errors.  Column 10 gives the number $N$ of field stars used to determine the systemic velocity, and Columns 11 -- 14 give the RA and Dec field systemic velocities with their errors. The total residual transverse velocity $v_\perp$ and error are given in Columns~15 and 16, while $v_r \sin i$ and mass are respectively listed in Columns 17 and 18.

\clearpage
\centerwidetable
\begin{rotatetable*}
\begin{deluxetable*}{lcccccccccccccccccc}
\tabletypesize{\footnotesize}
\tablewidth{0pt}
\tablecaption{Kinematic Data for SMC Field OB and OBe Stars \label{tab:Table 1}}
\tablehead{
\colhead{ID\tablenotemark{a}} & \colhead{Class\tablenotemark{b}} & \colhead{SpT} & \colhead{$\alpha$ (J2016)} & \colhead{$\delta$ (J2016)} & \colhead{Target $v_{\alpha}$} & \colhead{err} & \colhead{Target $v_\delta$} & \colhead{err} & \colhead{N} & \colhead{Field $v_{\alpha}$} & \colhead{err} & \colhead{Field $v_{\delta}$} & \colhead{err}  & \colhead{$v_\perp$} & \colhead{err} & \colhead{$v_r \sin i$} & \colhead{Mass} \\
&  &  & {(deg)} & {(deg)} & {($\kms$)} &  & {($\kms$)} &  & & {($\kms$)} &  & {($\kms$)} & & {($\kms$)} & {$\kms$} & {$\kms$} & {$M_\odot$}
}
\startdata
107 & -,-,-,e & Be3 & 10.11696 & -73.54262 & 151 & 6 & --379 & 6 & 55 & 171 & 6 & --365 & 0 & 25 & 11 & -- & 14.9 \\
298* & -,-,-,e & B1e3+ & 10.18283 & -73.40645 & 162 & 8 & --372 & 8 & 74 & 168 & 5 & --368 & 1 & 7 & 13 & -- & 17.0 \\
1037 & -,-,-,- & B0.5V & 10.38770 & -73.42581 & 129 & 10 & --358 & 10 & 61 & 177 & 6 & --371 & 9 & 50 & 17 & 92 & 14.6 \\
1600 & E,-,-,- & O8.5V & 10.54135 & -73.23245 & 192 & 9 & --352 & 9 & 56 & 181 & 5 & --360 & 2 & 14 & 14 & 91 & 26.8 \\
1631 & -,-,-,e & B1e2 & 10.55134 & -73.38681 & 151 & 9 & --340 & 8 & 55 & 164 & 8 & --371 & 7 & 33 & 16 & 197 & 14.6 \\
1830 & -,-,-,- & B0.5III & 10.60000 & -73.28109 & 134 & 6 & --359 & 6 & 63 & 171 & 5 & --371 & 3 & 40 & 10 & 86 & 22.3 \\
1952* & -,-,-,e & B1e2 & 10.63259 & -73.36706 & 141 & 10 & --346 & 9 & 63 & 175 & 6 & --356 & 3 & 36 & 15 & -- & 18.0 \\
2034 & -,-,-,e & Be & 10.64990 & -73.79843 & 177 & 7 & --390 & 7 & 33 & 186 & 7 & --368 & 4 & 23 & 13 & -- & 17.2 \\
2093 & -,-,-,e & B1e3+ & 10.66719 & -73.49800 & 146 & 7 & --331 & 7 & 83 & 177 & 5 & --366 & 1 & 46 & 12 & -- & 13.1 \\
2666* & -,-,-,e & B1.5e3+ & 10.79250 & -73.57531 & 182 & 9 & --349 & 9 & 59 & 181 & 7 & --360 & 0 & 11 & 14 & -- & 14.0 \\
\enddata
\tablenotetext{a}{IDs are from \cite{Massey02}. Stars with an asterisk (*) next to their ID are RIOTS4 data that were not included in Paper II. Stars with a plus sign (+) next to their ID have spectral types taken from their quiescent phase, but are identified as emission-line stars by \cite{Lamb16} or SIMBAD.}
\tablenotetext{b}{E = Eclipsing binary, S = Double-lined spectroscopic binary, X = High-mass X-ray binary, e = OBe star.}
\tablecomments{Table \ref{tab:Table 1} is published in its entirety in the machine-readable format. A portion is shown here for guidance regarding its form and content.}
\end{deluxetable*}
\end{rotatetable*}
\clearpage
\section{Updated Gaia DR3 Kinematics} \label{sec:Results}

Paper~II adopted a threshold runaway velocity of $v_\perp$ $\geq 30\ \kms$, which is equivalent to a 3-D space velocity $\geq 37\ \kms$, due in part to the high median error of $27\ \kms$ in that work. Now that we have substantially reduced this error to 16 $\kms$, we adopt the conventional 3-D space velocity runaway definition of 30 $\kms$, which corresponds to a transverse  $v_\perp\geq 24\ \kms$. We define stars that are unbound from their clusters with proper motions below this threshold to be walkaway stars. Our total OB field sample consists of 201 runaways, or 60\% $\pm$  5\% of our sample, and the remaining 135 stars are presumed to be walkaways, or 40\% $\pm$ 4\% of our sample. However, the walkaway sample is significantly underestimated since many slow-moving stars have not had time to reach the field yet, as discussed in Paper~II. Such unbound stars that are still close to their parent clusters would be dominated by walkaways, and they would not be in our field star sample.  

\begin{figure}
\begin{center}
    \includegraphics[scale=0.4] {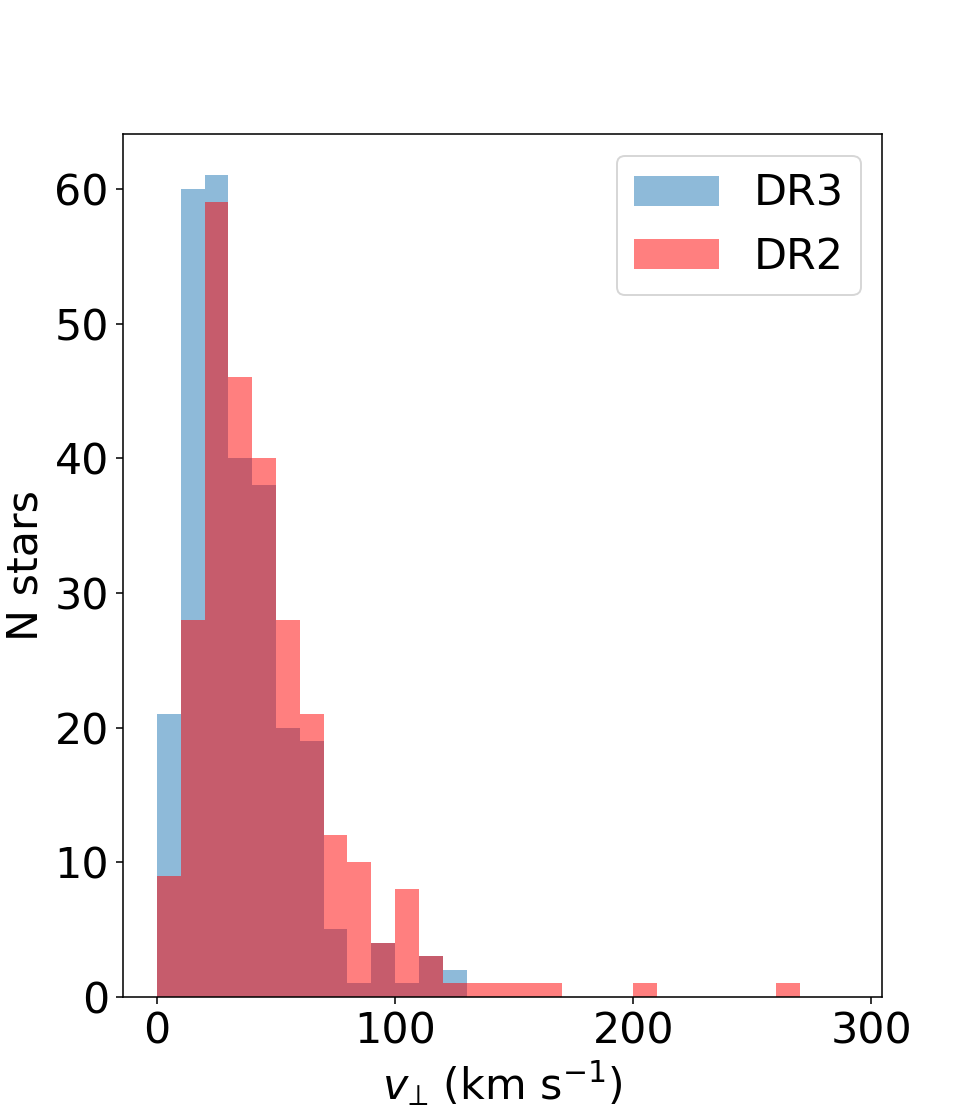}
\end{center}
\caption{The sample velocity distributions, comparing Gaia DR2 velocities from Paper I and Gaia DR3 velocities from this paper; we plot the 275 objects contained in both samples. \label{fig:Figure 2}}
\end{figure}

We find that, compared to the DR2 measurements in Paper I,  our median transverse velocity decreases by $\sim35$\%, from 39 $\kms$ to 29 $\kms$, showing that the velocities used in Papers I and II were systematically overestimated. 
This is to be expected, since astrometric errors almost always generate larger measured values. As seen in Figure \ref{fig:Figure 2}, our previous DR2 velocity distribution extends out to speeds greater than 150 $\kms$, with a maximum of 260 $\kms$. This high-velocity tail is greatly diminished in our new velocity distribution, now with speeds only reaching $\sim 130\ \kms$. The median DR3/DR2 velocity ratio is 0.77 with a standard deviation of 0.25. However, the revised velocities can be much smaller for given individual objects.  For instance, we obtained a particularly high velocity of 90 $\pm$ 30 $\kms$ for the well-known HMXB SMC X-1 (M2002-77458) discussed in Paper II, but now we measure it to be traveling at only 27 $\pm$ 12 $\kms$. Our tentative interpretation of its apparently unusually high velocity in that work therefore should be revised accordingly.
\begin{figure*}
\begin{center}
    \includegraphics[scale=0.4] {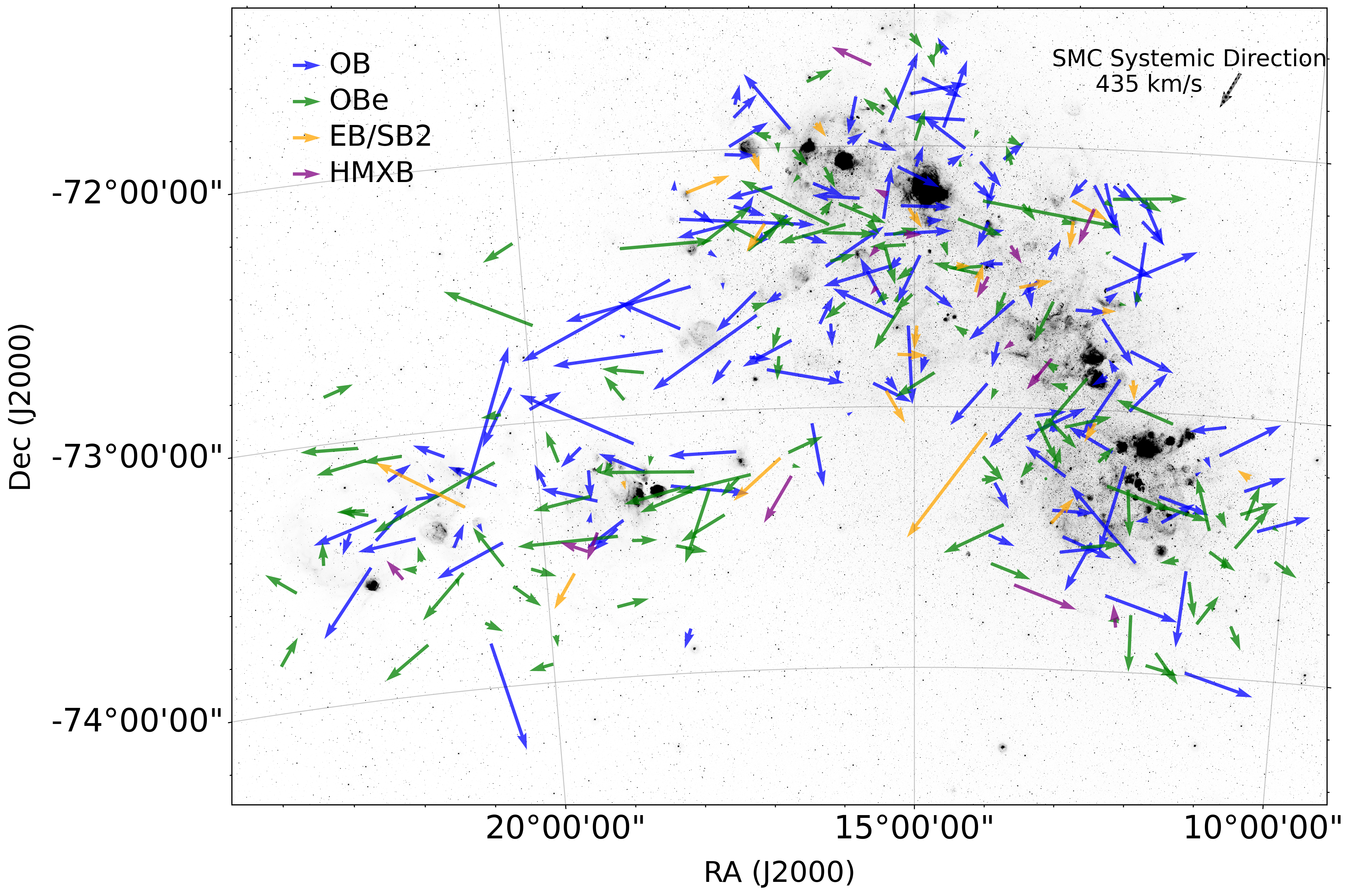}
\end{center}
\caption{Proper motion vectors for our sample stars, with subpopulations color-coded as shown. The length of the legend vectors in the top left corresponds to  25 $\kms$; the SMC systemic velocity vector in the top right is not drawn to scale. \label{fig:Figure 3}}
\end{figure*}

\begin{figure*}
\begin{center}
    \includegraphics[scale=0.4] {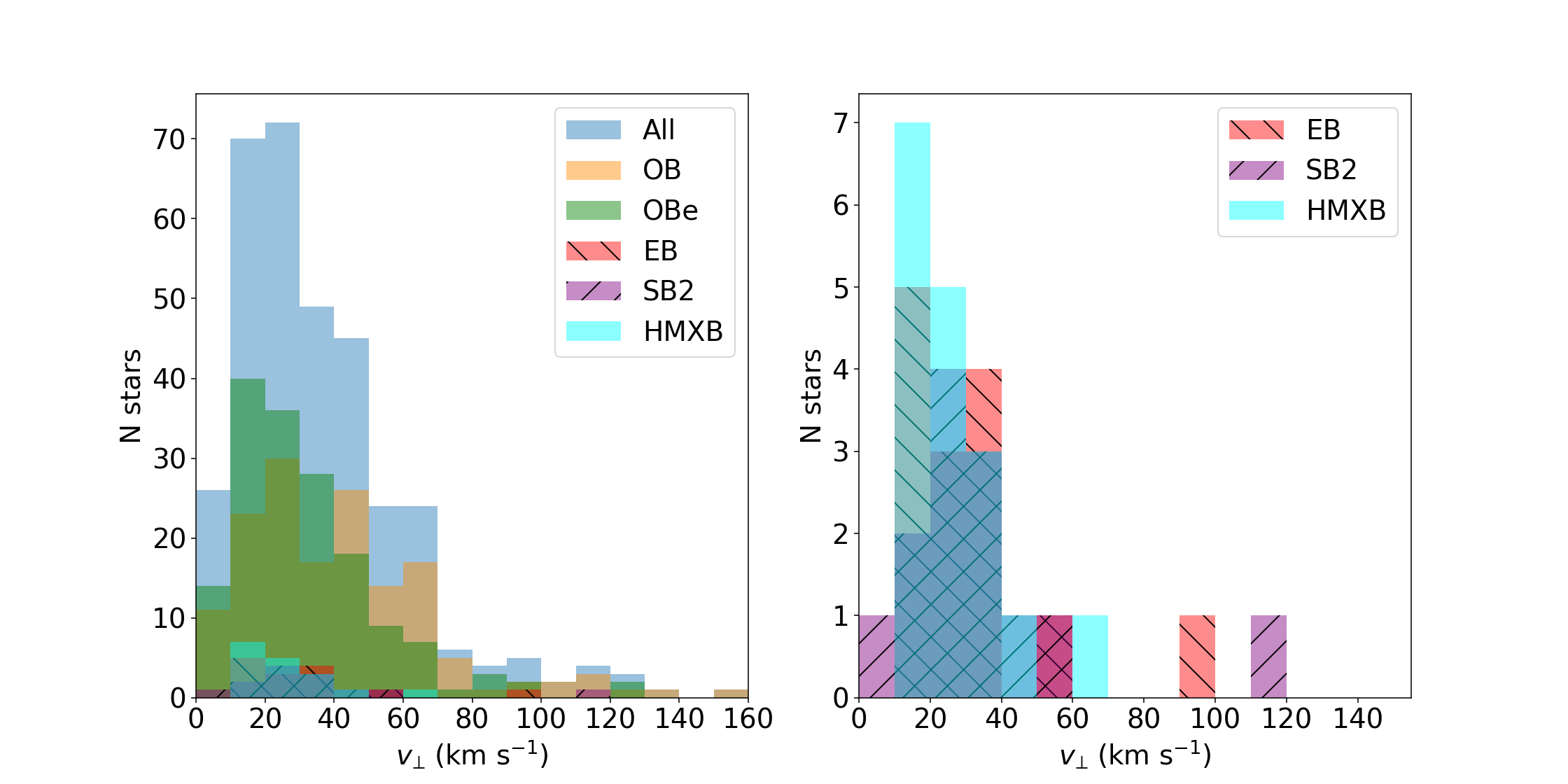}
\end{center}
\caption{Transverse velocity distributions for different subpopulations, color-coded as shown. There are 336 total stars, including 160 OBe stars, 17 HMXBs, 14 EBs, 13 SB2s, and 154 unclassified OB stars that do not belong to any of these groups.  
The right panel is a zoom on the shown subpopulations.
\label{fig:Figure 4}}
\end{figure*}

\begin{deluxetable}{lcc}
\tabletypesize{\footnotesize}
\tablewidth{0pt}
\tablecaption{Subpopulation kinematics
\label{tab:Table 2}}
\tablehead{
\colhead{Subgroup} & \colhead{Sample size} & \colhead{Median $v_\perp$} \\
 & & (km s$^{-1}$)
}
\startdata 
All stars & 336 & 29 \\
OB & 154 & 35 \\
OBe\tablenotemark{a} & 160 & 25 \\
HMXB\tablenotemark{b} & 17 & 22 \\
EB\tablenotemark{c} & 14 & 26 \\
SB2\tablenotemark{c} & 13 & 26 \\
\hline
Total DES objects & 177 & 33 \\
Total BSS objects & 159 & 25\\
\hline
OB fast rotators\tablenotemark{d} & 42 & 30 \\
OB slow rotators\tablenotemark{d} & 112 & 37 \\
\enddata
\tablenotetext{a}{Includes 142 apparently single OBe stars, 16 emission-line HMXBs, and 2 OBe non-compact binaries.}
\tablenotetext{b}{Includes 16 emission-line HMXBs and 1 non-emission HMXB.}
\tablenotetext{c}{These include 10 EBs, 9 SB2s, and 4 systems that are both EBs and SB2s. The 2 OBe non-compact binaries (1 EB and 1 EB+SB2) are also included.}
\tablenotetext{d}{OB-star fast and slow rotators have $v_r\sin i >$ and $< 150\ \kms$, respectively.}
\end{deluxetable}

Figure \ref{fig:Figure 3} shows a vector map of our sample, identifying classical OBe stars, HMXBs, EBs/SB2s, and the remaining normal OB stars. The velocity distributions of these subpopulations are shown in Figure \ref{fig:Figure 4}. This includes the velocity distribution of OBe stars, which was not explicitly shown in Paper II.  We note that individual stars may be members of multiple subpopulations in Figure~\ref{fig:Figure 4}; for instance, the OBe group contains almost all HMXB stars, the EB group contains EB+SB2 stars, and so on.  Additionally, we present the median velocity of each subgroup in Table \ref{tab:Table 2}. 

The BSS produces fewer high-velocity runaways than the DES \cite[]{PeretsSubr,Renzo19}. Paper~I found that the velocity distribution and median velocity of HMXBs are slower than those of EBs and SB2s, supporting this scenario. Paper~II showed that since OBe stars are believed to be spun up from mass transfer in binary systems, the BSS appears to be the predominant mechanism by which these stars are ejected. This connection is supported by the fact that only two of our pre-SN binary systems, [M2002] SMC-24119 and 30744, are OBe stars. The former is a new EB identification. Conversely, we assume that normal field OB stars are likely dominated by the DES mechanism, a premise further supported by the statistics and kinematics of these populations in Paper~II.  

As seen in Figure \ref{fig:Figure 4}, the normal OB distribution (orange histogram) has a high-velocity tail that extends out to $\sim$160 $\kms$. On the other hand, the OBe distribution (green histogram) only extends out to  $\sim$80 $\kms$. Additionally, the median OB star velocity is 35 $\kms$ compared to 25 $\kms$ for the OBe stars, as seen in Table \ref{tab:Table 2}. These data are consistent with the expectation that BSS-ejected stars should generally be moving more slowly than the DES-ejected stars. 

\begin{figure*}
    \includegraphics[width=0.5\textwidth]{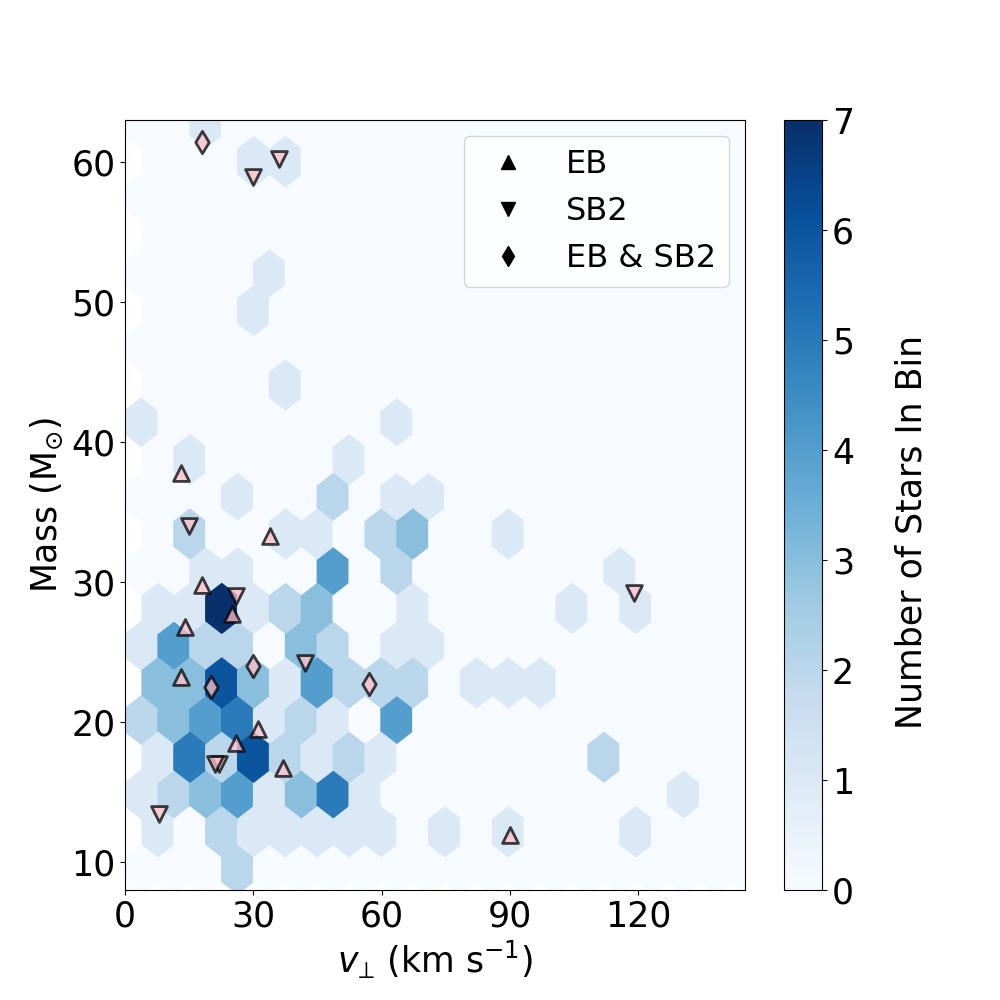}%
    \includegraphics[width=0.5\textwidth]{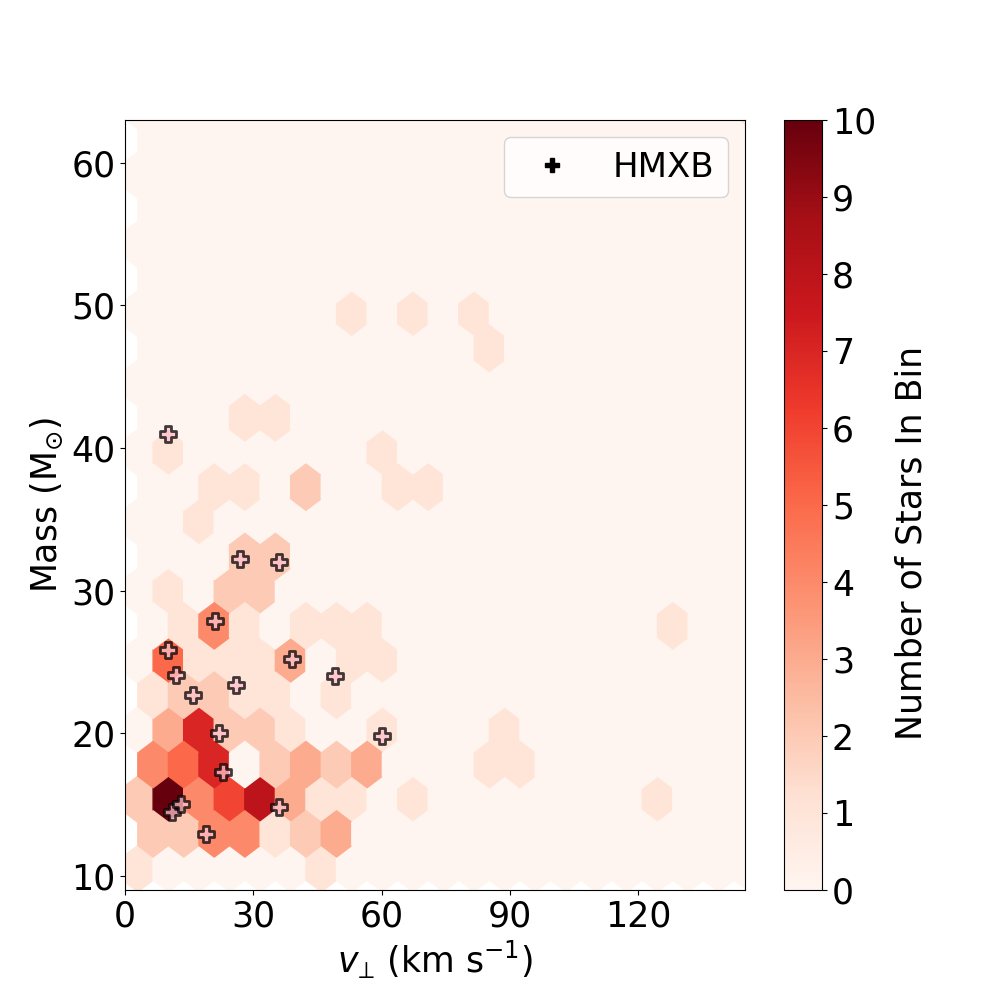}%
    \caption{Mass-velocity hexabin density plots for our DES stars (blue) and BSS stars (red). The legends define the symbols for the overplotted tracer populations. }
    \label{fig:Figure 7}
\end{figure*}

As discussed in Section 1, we expect the DES to produce not only faster, but also more massive runaway stars compared to the BSS. Moreover, studies predict that the velocities of SN-ejected systems should decrease with mass while those of dynamically ejected stars should increase with mass \cite[]{PeretsSubr,OhKroupa,Renzo19}. This behavior can directly be seen in Figure \ref{fig:Figure 7}, which shows the 2-D distribution of mass vs velocity for the 159 BSS stars and 177 DES stars with available masses (Table~\ref{tab:Table 1}). 

In the left panel of Figure~\ref{fig:Figure 7}, for the dynamically ejected systems, 
there may be evidence for bimodality in the kinematics, which suggest a low-mass, low-velocity population, and also a high-mass, high-velocity component, with an apparent minimum around (32 $\kms$, 26 M$_\odot$) between these groups. 
The existence of this bimodality appears to be robust to algorithms for local systemic velocity determinations, and
we tentatively suggest that it may be real. 
Dynamical ejections are expected to include different processes that affect the kinematics, such as
single-star versus binary system ejections, single-star versus binary or multiple-star interactions, and mergers.  Alternatively, some objects in the fast component may be two-step ejections that do not form decretion disks, and therefore do not appear as OBe stars.  However, we note that this last scenario is not supported by the rotation velocities (see Section~\ref{sec:vsini} below).

In the right panel of Figure \ref{fig:Figure 7}, for our BSS stars, we observe a single population of stars centered around $\sim20$ $\kms$ and 18\ M$_\odot$. These lower values compared to the DES are consistent with expectation since it is difficult for the BSS to generate high mass, high-velocity stars \citep[e.g.,][]{Renzo19}. However, we do see a smaller population of stars moving at much higher speeds than expected for SN acceleration alone; very few BSS objects are expected at runaway velocities $> 30\ \kms$ \citep[e.g.,][]{Renzo19}. As suggested in Paper II, these may be two-step ejections, meaning that they have undergone both a dynamical ejection and a subsequent SN reacceleration \citep{Pflamm-Altenburg10}. The frequency of two-step ejections will be examined
in the next section. 

Turning to our direct tracer populations, our EBs and SB2s are pre-SN systems, and thus they must have been ejected into the field dynamically. As seen in the right panel of Figure \ref{fig:Figure 4}, we observe a high-velocity tail for our non-compact binaries.  Our fastest EB, [M2002] SMC-81258, has a transverse velocity of 90 $\pm$ 24 $\kms$; and our fastest SB2, [M2002] SMC-36213, has a transverse velocity of 119 $\pm$ 14 $\kms$, both of which we inferred to be moving faster in Paper II. However, these speeds still 
suggest
a DES origin. Out of our 23 total non-compact binaries, over half (13) are runaway systems. Comparing our non-compact binary distributions with our normal OB star velocity distribution, we see that they both possess similar high-velocity tails, as expected if both originate from the DES. 

However, a two-sample Kolmogorov-Smirnov (K-S) test comparing our normal OB stars and non-compact binaries results in a $p$-value 0.03, and the median EB and SB2 velocities are significantly lower than the median OB star velocity (Table~\ref{tab:Table 2}). We also see that the distribution of the non-compact binaries in the DES mass-velocity plot in Figure~\ref{fig:Figure 7} does not show strong evidence of the bimodal distribution identified earlier for the OB stars.  Instead, the non-compact binaries appear to preferentially associate with the lower-mass, lower-velocity kinematic component. This may be reasonable since dynamical ejections of binary systems generate slower ejections than for single stars.
However, there are some exceptions, and
it is interesting to note that  
both the most massive binaries, and also the fastest binaries, 
are among the most extreme objects in the OB population (Figure~\ref{fig:Figure 7}), and therefore
belong to the high-mass, high-velocity DES component suggested above.

Conversely, the HMXBs are post-SN systems, and they must have experienced the BSS mechanism. As seen in the right panel of Figure \ref{fig:Figure 4}, all of our HMXBs show Gaia DR3 velocities that are now confined to $v_\perp$ $<$ 60 $\kms$. In addition, our median HMXB velocity has decreased from 28 $\kms$ in Paper II to 22 $\kms$ here. This better agrees with the work of \cite{Renzo19}, who predict a median systemic velocity of 20 $\kms$ for neutron star binaries with main sequence companions. Since we consider only field stars, the overall median HMXB velocity may be even lower than this prediction. We also reconfirm the similarity in the kinematics of our OBe stars and HMXBs (Paper II) seen in Figures~\ref{fig:Figure 4} and \ref{fig:Figure 7}; both velocity distributions peak at $\sim$ 10 $\kms$ and then steadily decline, and the populations compare well in the mass-velocity parameter space shown in Figure~\ref{fig:Figure 7}.

A K-S test comparing our OBe stars and HMXBs yields a $p$-value of 0.56, consistent with these populations having a similar origin.  The fact that the statistical probability is not even higher is likely due to the fact that
our OBe stars have a high-velocity tail, unlike the HMXBs.  This difference may be due in part to small number statistics for the HMXBs, but as noted earlier, the OBe population also may be supplemented by systems ejected by the two-step mechanism.

\subsection{The nature of OB fast rotators}\label{sec:vsini}

There is a possibility that our BSS sample is incomplete due to additional, unidentified BSS stars that are not OBe stars. In particular, normal OB stars that are fast rotators may also originate from binary mass transfer and such field stars may therefore be BSS products. In total, 42 out of our 154 normal OB stars are fast rotators, leaving 112 slow-rotator OB stars, and 159 OBe stars (Table~\ref{tab:Table 2}). Thus, if the OB fast rotators are all BSS products, then 
BSS ejections
would dominate our sample, which is currently not the case, at face value.

\begin{figure}
\begin{center}
    \includegraphics[scale=0.3] {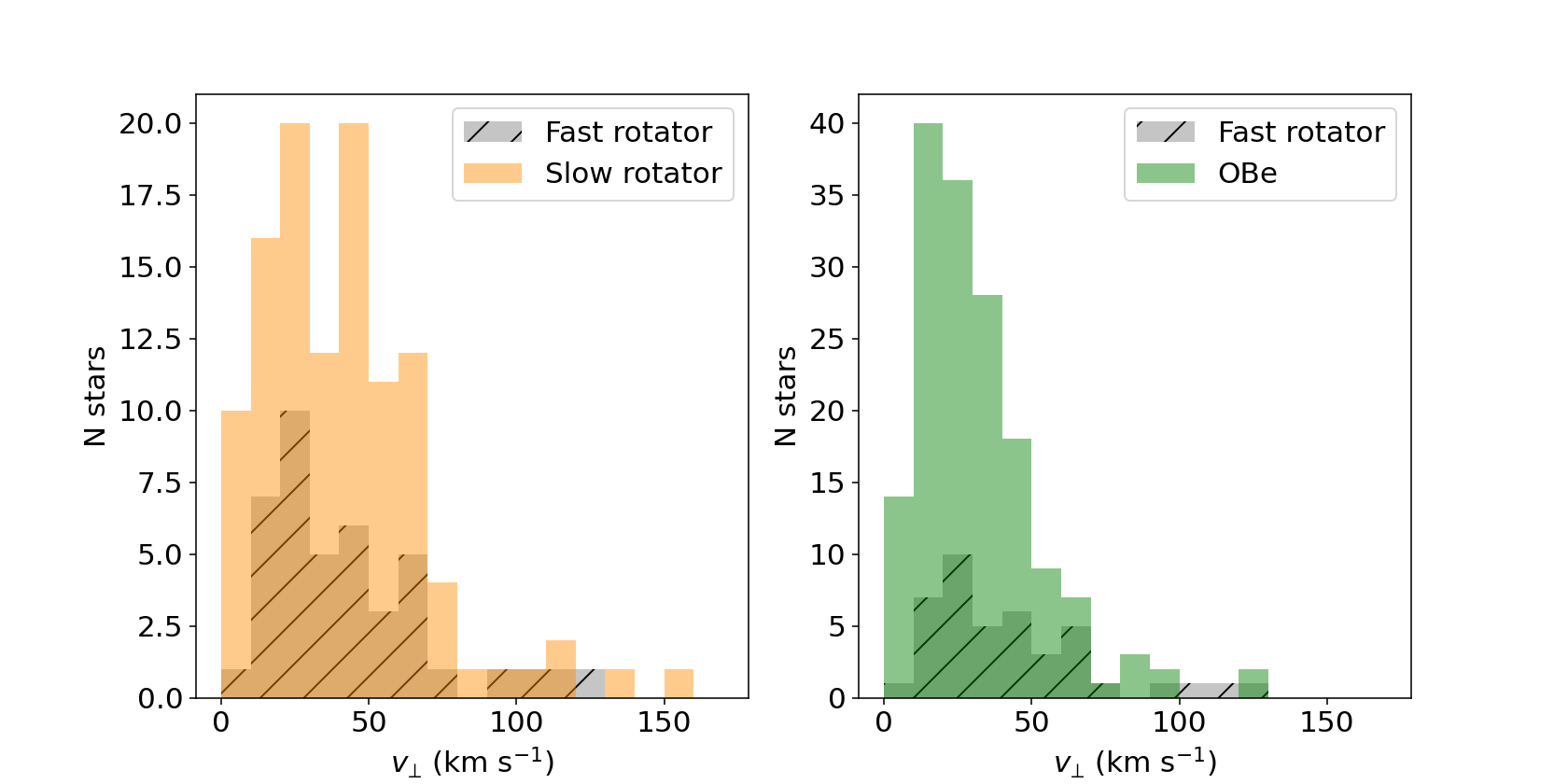}
\end{center}
\caption{Comparison of velocity distributions for fast-rotator OB stars with the remaining (slow-rotator) OB stars (left panel), and OBe stars (right panel). \label{fig:Figure 5}}
\end{figure}

\begin{figure*}
    \includegraphics[width=0.5\textwidth]{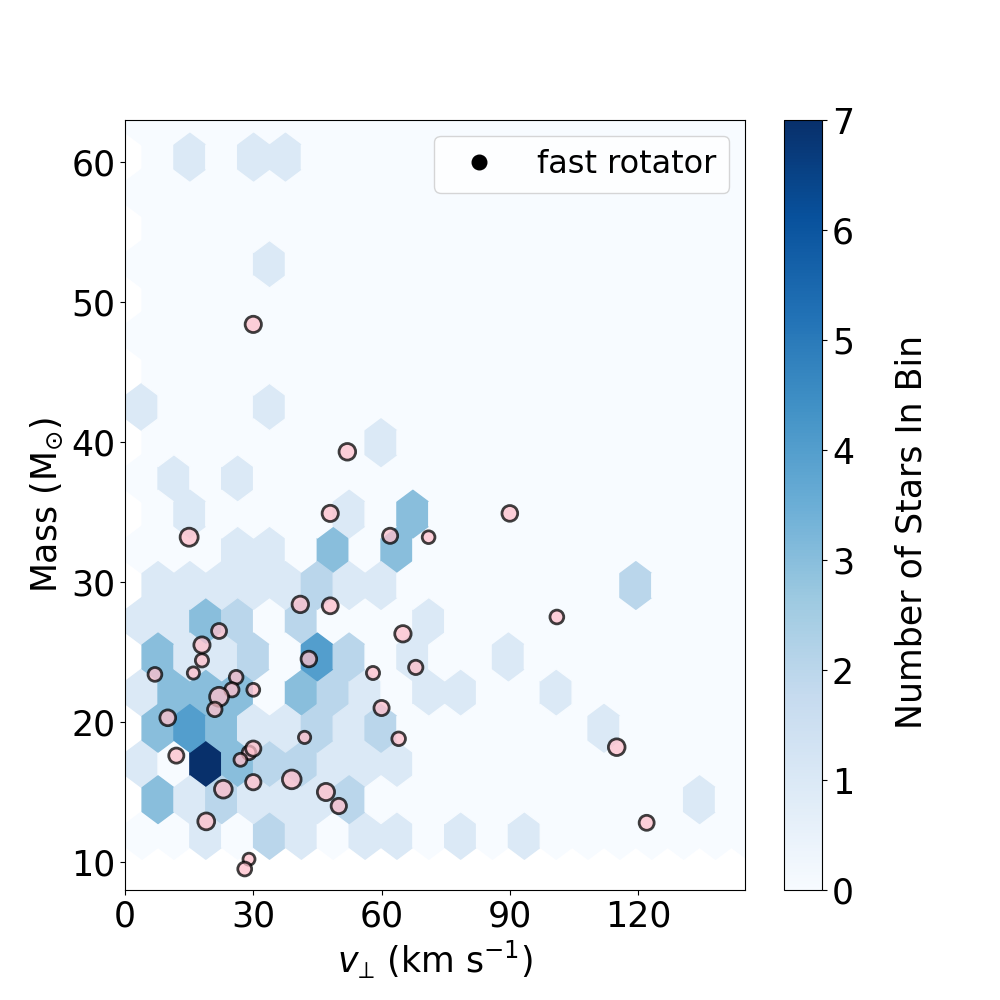}%
    \includegraphics[width=0.5\textwidth]{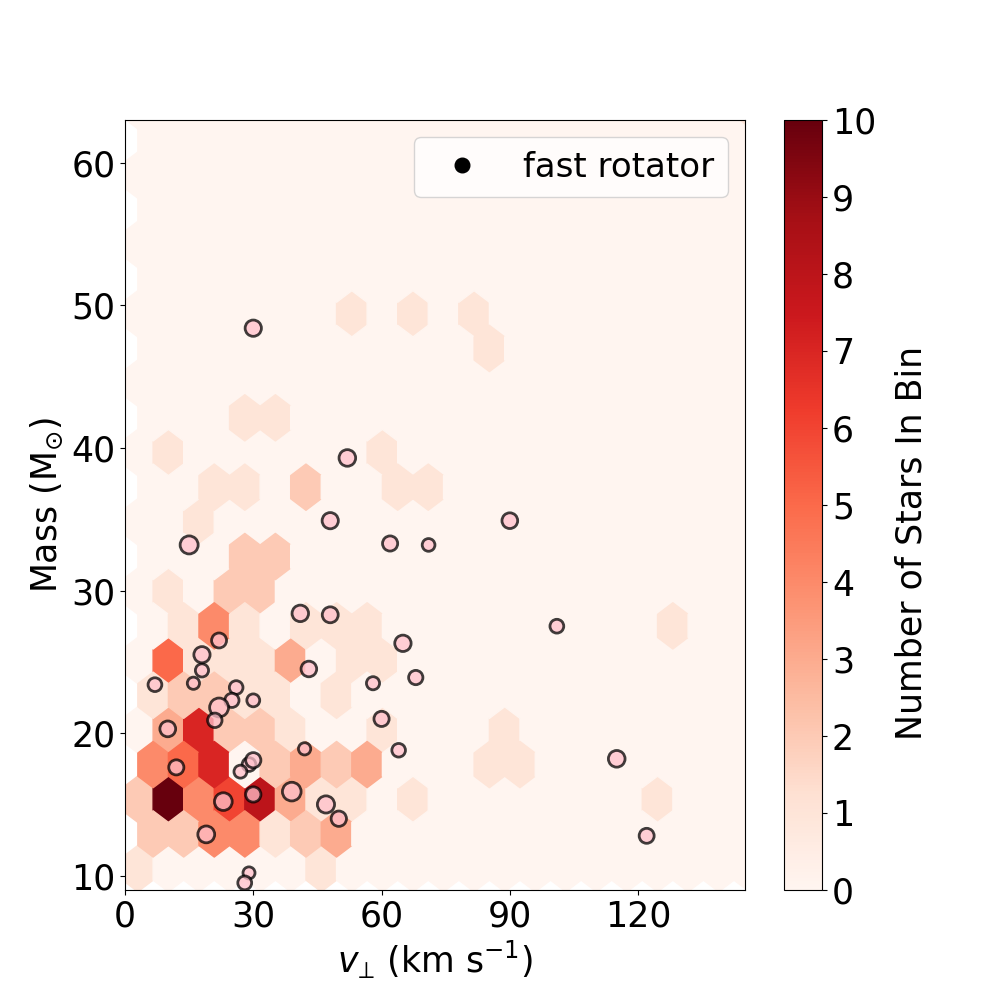}%
    \caption{Mass-velocity hexabin density plots for our DES stars (blue) and BSS stars (red); the OB fast rotators are excluded from the density bins and are instead overplotted as circles whose sizes are proportional to $v_r\sin i$. }
    \label{fig:Figure 6}
\end{figure*}

In Figure \ref{fig:Figure 5}, we plot the velocity distribution of OB stars with $v_{r}\sin i > 150$ $\kms$ against the remaining OB and OBe star distributions. 
Surprisingly, Figure \ref{fig:Figure 5} shows that the kinematics of our fast-rotator OB stars behave more similarly to those of the slow-rotator OB stars than the OBe stars; both slow- and fast-rotator OB star velocity
distributions share a bimodality with peaks at $\sim$25 $\kms$ and $\sim$40 $\kms$. 
In addition, a K-S test between our fast and slow-rotator OB stars gives a $p$-value of 0.92, strongly supporting a scenario that the two samples come from the same parent distribution. On the other hand, our OBe star distribution peaks at $\sim$20 $\kms$ and then steadily decreases with increasing velocity, unlike our fast rotators. A K-S test between the fast-rotator OB and OBe samples gives a $p$-value of 0.09, confirming that the two samples very likely do not come from the same distribution. The mass-velocity plots in Figure~\ref{fig:Figure 6} also appears to confirm that the normal OB stars do not differentiate by $v_r\sin i$.
These results indicate that {\it we cannot conclude that the OB fast rotators are BSS objects.}  The origin of their high $v_r\sin i$ appears to have a different origin than for OBe stars; one possibility is that they may be merger products \citep[e.g.,][]{Leonard1995}. We therefore continue to assume, as done in Paper~II, that the fast rotator OB stars belong with the other normal OB stars and are dominated by the DES ejection mechanism.

\section{DES vs BSS Ejections} \label{sec:model}

Thus, the kinematics of OB and OBe stars remain generally compatible with interpreting these populations as being dominated by DES and BSS ejections, respectively.  We will proceed with this assumption, although there are caveats which will be discussed below.

Our observed statistics allow us to estimate the relative frequencies of both ejection mechanisms and compare them to those produced by models. Out of our sample of 336 field stars, 177 are normal OB stars or non-compact binaries, and thus putative DES objects, and 159 are OBe stars or HMXBs (Table~\ref{tab:Table 2}), and thus presumed BSS objects. This corresponds to a total DES/BSS ratio among our field stars of 1.11 $\pm$ 0.12. For only the runaway stars with $v_\perp>$ 24 $\kms$, the ratio increases to 1.45 $\pm$ 0.21.

As discussed in Paper II, walkaways are highly incomplete due to the RIOTS4 survey selection criterion for deep field objects, which are specified to be at least 28 pc away from any other OB candidate. Thus, most walkaways do not have the necessary speeds to travel that far from the time they were ejected. Paper II determined that the correction factor for the walkaway selection bias is $\sim$ 2.4. As argued in that work, a similar correction factor for runaways is considered relatively unimportant in comparison.  We therefore focus on the runaway DES/BSS ratio.
We can also consider the implied walkaway to runaway ratios (W/R) for the two ejection mechanisms, obtaining the ratios shown in Table~\ref{tab:Table 3}.

\begin{deluxetable*}{lcccc}
\tablecaption{Numbers and Ratios of Walkaway and Runaway Field OB Stars
\label{tab:Table 3}}
\tabletypesize{\footnotesize}
\tablewidth{0pt}
\tablehead{
\colhead{} & 
\colhead{Numerator\tablenotemark{a}} & \colhead{Denominator} &
\colhead{Ratio}  & \colhead{Model\tablenotemark{b}}}
\startdata 
Runaway DES/BSS & 119 & 82 & 1.5 $\pm$ 0.2 & 1.7 \\
Total W/R & 324 & 201 & 1.6 $\pm$ 0.1 & 1.8 \\
BSS W/R & 185 & 82 & 2.3 $\pm$ 0.3 & 2.3 \\
DES W/R & 139 & 119 & 1.2 $\pm$ 0.1 & 1.5 \\
\enddata
\tablenotetext{a}{Numbers of walkaways are corrected by a factor of 2.4 for incompleteness (see Paper II).}
\tablenotetext{b}{See text and Table~\ref{tab:Table 4} for details about the model predictions.}
\end{deluxetable*}
We use the model developed in Paper II, which combines independent DES and BSS studies from \cite{OhKroupa} and \cite{Renzo19}, respectively, to estimate the expected kinematics produced by ejections from a singular stellar population.  We refer the reader to Paper~II for complete details.  The key parameters for the DES model are: the W/R ratio, the total ejection fraction $f_{\rm ej}$, and the non-compact binary ejection fraction $f_{\rm bin}$. The Oh \& Kroupa models are based on evaluating the system at an age of 3 Myr, and thus applying this model to the entire SMC population assumes that the properties at this age are similar to those of an age-averaged population of O stars.  We caution that early B stars in our sample can live to ages $\gtrsim 3\times$ longer, and thus the models may be biased toward effects dominated by more massive stars, which tend to be ejected earlier.
\begin{deluxetable}{lcc}
\tablecaption{Model Runaway and Walkaway Frequencies\tablenotemark{a} \label{tab:Table 4}}
\tablehead{\colhead{} &\colhead{Runaways} & \colhead{Walkaways}} 
\startdata
DES & &  \\
\quad All Ejected & 0.08 & 0.12 \\
\quad {\bf Pre-SN} & {\bf 0.06} & {\bf 0.09} \\
\quad Post-SN & 0.019 & 0.029 \\
\quad Pre-SN, binaries & 0.049 & 0.073 \\
\quad Post-SN, binaries & 0.017 & 0.026 \\
\quad Two-step & 0.020 & 0.018 \\
\hline
BSS & &  \\
\quad Unbound & 0.012 & 0.021 \\
\quad Bound & 0.0080 & 0.060 \\
\quad All Ejected & 0.020 & 0.080 \\
\quad Total pure BSS & 0.016 & 0.064 \\
\quad {\bf With two-step} & {\bf 0.036} & {\bf 0.083} \\
\hline
Total Predicted & {\bf 0.10} & {\bf 0.17}  \\
{\it Total Observed} & 0.10 & 0.15  \\
\hline
DES/BSS ratio & {\bf 1.7} & 1.1 \\
Pre-SN binaries, sub-pop & {\bf 0.50} & {\bf 0.42} \\
\hline
\enddata
\tablenotetext{a}{Population frequencies within an entire OB population, with
adopted parameters: (1) DES walkaway:runaway branching ratio W/R of 60:40, (2) DES ejection fraction of 0.2, (3) DES binary frequency of 0.8, and (4) fraction of 0.2 post-SN walkaways that are reaccelerated to runaways. Values for comparison to observations are boldface.}
\end{deluxetable}

The main key parameter for the BSS model is the fractions of BSS walkaways that are ejected from their birth clusters, for both single stars and those still bound to compact remnants.  In Paper~II, we adopted a value of 50\% for both bound and unbound objects, but here we revise the value for unbound stars to 20\%.  
The revised value is more consistent with the models of \citet{Renzo19}, which show that unbound objects have slower velocities; however, we note that this adjustment does not strongly affect results.
Another free parameter is the fraction $f_{\rm 2step}$ of DES-ejected, non-compact binaries that result in two-step ejections that are runaways. 

We update the model in Paper~II to better fit the revised kinematics presented here.  We find a reasonable match for parameters corresponding to W/R = 60:40,  $f_{\rm ej}$ = 0.2, $f_{\rm bin}$ = 0.8, and $f_{\rm 2step} = 0.2$. Tables \ref{tab:Table 3} and \ref{tab:Table 4} show the resulting DES and BSS populations with comparisons to observed values. Following Paper~II, Table~\ref{tab:Table 4} shows total frequencies of the subpopulations for an SMC field star population corresponding to 25\% of all OB stars.  Our updated velocities imply runaway and walkaway frequencies for this total population of $39\%\pm 5\%$  and $61\%\pm 4\%,$ respectively.  Thus, consistent with expectation, walkaways now dominate over runaways, which was not the case in Paper~II. We still find that the two-step mechanism may dominate BSS ejections for runaway stars, but not for walkaways (Table~\ref{tab:Table 4}). For comparison, parameters for the best model in Paper~II were W/R = 50:50, $f_{\rm ej}$ = 0.3 $f_{\rm bin}$ = 1.0, and $f_{\rm 2step} = 0.33$.  

The DES runaway fractions in the models by \citet{OhKroupa} are generally on the order of 0.1 -- 0.2 rather than 0.4 -- 0.5 as suggested by our results.  This is a bit unexpected, since as mentioned above, age effects imply that our observations 
could be biased toward slightly lower-mass stars than the DES models by \citet{OhKroupa}; since more massive stars tend to have larger runaway velocities, the bias should cause the observed $f_{\rm run}$ to be lower, not higher, than the DES models. However, we found earlier that the DR2 proper motions in Paper~II were systemically overestimated, so it is possible that a similar effect is still happening here with our DR3 data.  If so, then our runaway fraction remains overestimated. Another possibility is that our adopted correction factor of $\sim 2.4$ from Paper~II is too low. 

It is also important to note that our identification of OB stars with DES and OBe stars with BSS is likely overly simplistic.  There may be SN-ejected objects that are not OBe stars, and some OBe stars may have circumstellar disks for reasons other than binary mass transfer, and so may not be accelerated by SNe.  There are also a few OBe stars that may be dynamically ejected and pre-SN, unidentified binaries. Additionally, our non-compact binary sample of EBs and SB2s is underestimated since there are likely additional binaries that have not been identified yet (Vargas-Salazar et al., in preparation); such systems are direct tracers of the DES mechanism. 

There are also substantial theoretical uncertainties. Dynamical collisions can cause partner exchanges and mergers that can significantly affect a given binary system's subsequent evolution.  Such encounters are relatively rare; \citet{Ivanova2005} show that for a typical $10^5\ M_\odot$ cluster, the rate
of
evolution-induced binary destruction within the first 20 Myr is only $\sim2\times$
higher than without accounting for dynamical interactions.  Thus, to first order, we can treat DES and BSS as independent mechanisms, but a correct treatment will need an integrated approach.  There are also substantial uncertainties in supernova kicks \citep[e.g.,][]{Renzo19}, fallback, and neutron star vs black hole progenitors \citep[e.g.,][]{OConnor2011, Sukhbold2016}, all of which can substantially affect BSS velocities. Thus, a close match between this simple, single-population model and the observations is not necessarily expected. But the fact that the agreement is quite good does suggest that both our interpretation of the data and the general model parameters are reasonable.

\section{Conclusion} \label{sec:Conclusion}

In this work, we present updated proper motions for the field OB and OBe stars in the Small Magellanic Cloud, using data from Gaia DR3 to shed light on the stellar cluster ejection mechanisms. Our sample of 336 stars is based on the spatially complete RIOTS4 survey, for which we include some updated spectral classifications. 
We compare new velocities for this larger sample to those from our previous work based on Gaia DR2 \cite[]{DorigoJones2020}, and we find a decrease in median velocity from 39 $\kms$ to 29 $\kms$, confirming that our previous velocities were systematically overestimated due to observational errors. In addition, our median error decreased significantly from 27 $\kms$ to 11 $\kms$ due to improvements in determining the local systemic
velocities, in addition to more accurate Gaia proper motions. 

We show that the stellar kinematics are consistent with an interpretation that the population of ordinary OB stars is dominated by dynamically ejected stars (DES), while OBe stars are dominated by SN ejections (BSS).  This is supported by the fact that the kinematics of HMXBs, which are post-SN systems, are consistent with those of the OBe stars.  The non-compact binaries (EBs and SB2s), which are tracers of the DES, conversely show a high-velocity tail consistent with OB stars.  However, velocities of non-compact binaries are more strongly skewed to lower values than OB stars in general.  This may be attributed to slower ejection velocities for binary systems than for single stars.  The systematically slower OBe star velocity distribution is also consistent with expectations that BSS ejections are less energetic than DES ejections.  Thus, these data (Figures \ref{fig:Figure 4} and \ref{fig:Figure 7}) continue to support the scenario that OBe stars originate from mass transfer in close binary systems which spin up the mass gainers to speeds that generate the characteristic decretion disks of classical OBe stars.

We find that the kinematics of ordinary OB stars suggest a bimodality, with a slower, lower-mass component, and a faster, higher-mass component.  The origin of this bimodality remains to be determined, but it may be associated with binary vs single-star ejections; we find that the non-compact binary stars coincide much more strongly, but not exclusively, with the slower component.  

Fast-rotating, ordinary OB stars with $v\sin i> 150 \ \kms$ might have been expected to be similar in nature to OBe stars, but their proper motion velocity distributions are strongly and statistically consistent with those of the remaining, slow-rotating OB stars.  {\it Fast-rotator OB stars therefore do not appear to be dominated by BSS products.}  Their nature remains to be determined, but one possibility is that they are merger products \citep[e.g.,][]{Leonard1995}.

We compare our observations to predictions using the combined single stellar population model for DES ejections from \citet{OhKroupa} and BSS ejections from \citet{Renzo19} used by \citet{DorigoJones2020}.  We update the best-fit DES parameters to suggest a total ejection fraction of $\sim$ 0.2 and a non-compact binary ejection fraction of $\sim$ 0.8 for massive stars in the SMC. Given that the model is based on a single stellar population and that significant uncertainties in the observations remain, the generally good agreement simply suggests that both models and observations are reasonable, to first order.  

Our results still support the findings of Paper~II that DES ejections dominate over BSS products by a factor of $\sim1.7$ to generate runaways with space velocities $> 30\ \kms$.  The model assigns 2-step ejections to the BSS products; we still find that over half of all BSS runaways are 2-step ejections in this scenario (Section~\ref{sec:model}).

\begin{acknowledgments}
We thank Johnny Dorigo Jones for sharing his codes from Papers~I and II, and for early advice with coding.
We also thank the anonymous referee for comments that clarified our presentation.
\end{acknowledgments}

\bibliography{RunawaysIII}{}

\begin{thebibliography}{}
\expandafter\ifx\csname natexlab\endcsname\relax\def\natexlab#1{#1}\fi
\providecommand{\url}[1]{\href{#1}{#1}}
\providecommand{\dodoi}[1]{doi:~\href{http://doi.org/#1}{\nolinkurl{#1}}}
\providecommand{\doeprint}[1]{\href{http://ascl.net/#1}{\nolinkurl{http://ascl.net/#1}}}
\providecommand{\doarXiv}[1]{\href{https://arxiv.org/abs/#1}{\nolinkurl{https://arxiv.org/abs/#1}}}

\bibitem[{{Blaauw}(1961)}]{Blaauw61}
{Blaauw}, A. 1961, \bain, 15, 265

\bibitem[{{Dallas} {et~al.}(2022){Dallas}, {Oey}, \& {Castro}}]{Dallas2022}
{Dallas}, M.~M., {Oey}, M.~S., \& {Castro}, N. 2022, \apj, 936, 112, \dodoi{10.3847/1538-4357/ac8988}

\bibitem[{{Dorigo Jones} {et~al.}(2020){Dorigo Jones}, {Oey}, {Paggeot}, {Castro}, \& {Moe}}]{DorigoJones2020}
{Dorigo Jones}, J., {Oey}, M.~S., {Paggeot}, K., {Castro}, N., \& {Moe}, M. 2020, \apj, 903, 43, \dodoi{10.3847/1538-4357/abbc6b}

\bibitem[{{Fujii} \& {Portegies Zwart}(2011)}]{FujiiPZ}
{Fujii}, M.~S., \& {Portegies Zwart}, S. 2011, Science, 334, 1380, \dodoi{10.1126/science.1211927}

\bibitem[{{Gaia Collaboration} {et~al.}(2023){Gaia Collaboration}, {Vallenari}, {Brown}, {Prusti}, {de Bruijne}, {Arenou}, {Babusiaux}, {Biermann}, {Creevey}, {Ducourant}, {Evans}, {Eyer}, {Guerra}, {Hutton}, {Jordi}, {Klioner}, {Lammers}, {Lindegren}, {Luri}, {Mignard}, {Panem}, {Pourbaix}, {Randich}, {Sartoretti}, {Soubiran}, {Tanga}, {Walton}, {Bailer-Jones}, {Bastian}, {Drimmel}, {Jansen}, {Katz}, {Lattanzi}, {van Leeuwen}, {Bakker}, {Cacciari}, {Casta{\~n}eda}, {De Angeli}, {Fabricius}, {Fouesneau}, {Fr{\'e}mat}, {Galluccio}, {Guerrier}, {Heiter}, {Masana}, {Messineo}, {Mowlavi}, {Nicolas}, {Nienartowicz}, {Pailler}, {Panuzzo}, {Riclet}, {Roux}, {Seabroke}, {Sordo}, {Th{\'e}venin}, {Gracia-Abril}, {Portell}, {Teyssier}, {Altmann}, {Andrae}, {Audard}, {Bellas-Velidis}, {Benson}, {Berthier}, {Blomme}, {Burgess}, {Busonero}, {Busso}, {C{\'a}novas}, {Carry}, {Cellino}, {Cheek}, {Clementini}, {Damerdji}, {Davidson}, {de Teodoro}, {Nu{\~n}ez Campos}, {Delchambre}, {Dell'Oro}, {Esquej},
  {Fern{\'a}ndez-Hern{\'a}ndez}, {Fraile}, {Garabato}, {Garc{\'\i}a-Lario}, {Gosset}, {Haigron}, {Halbwachs}, {Hambly}, {Harrison}, {Hern{\'a}ndez}, {Hestroffer}, {Hodgkin}, {Holl}, {Jan{\ss}en}, {Jevardat de Fombelle}, {Jordan}, {Krone-Martins}, {Lanzafame}, {L{\"o}ffler}, {Marchal}, {Marrese}, {Moitinho}, {Muinonen}, {Osborne}, {Pancino}, {Pauwels}, {Recio-Blanco}, {Reyl{\'e}}, {Riello}, {Rimoldini}, {Roegiers}, {Rybizki}, {Sarro}, {Siopis}, {Smith}, {Sozzetti}, {Utrilla}, {van Leeuwen}, {Abbas}, {{\'A}brah{\'a}m}, {Abreu Aramburu}, {Aerts}, {Aguado}, {Ajaj}, {Aldea-Montero}, {Altavilla}, {{\'A}lvarez}, {Alves}, {Anders}, {Anderson}, {Anglada Varela}, {Antoja}, {Baines}, {Baker}, {Balaguer-N{\'u}{\~n}ez}, {Balbinot}, {Balog}, {Barache}, {Barbato}, {Barros}, {Barstow}, {Bartolom{\'e}}, {Bassilana}, {Bauchet}, {Becciani}, {Bellazzini}, {Berihuete}, {Bernet}, {Bertone}, {Bianchi}, {Binnenfeld}, {Blanco-Cuaresma}, {Blazere}, {Boch}, {Bombrun}, {Bossini}, {Bouquillon}, {Bragaglia}, {Bramante}, {Breedt},
  {Bressan}, {Brouillet}, {Brugaletta}, {Bucciarelli}, {Burlacu}, {Butkevich}, {Buzzi}, {Caffau}, {Cancelliere}, {Cantat-Gaudin}, {Carballo}, {Carlucci}, {Carnerero}, {Carrasco}, {Casamiquela}, {Castellani}, {Castro-Ginard}, {Chaoul}, {Charlot}, {Chemin}, {Chiaramida}, {Chiavassa}, {Chornay}, {Comoretto}, {Contursi}, {Cooper}, {Cornez}, {Cowell}, {Crifo}, {Cropper}, {Crosta}, {Crowley}, {Dafonte}, {Dapergolas}, {David}, {David}, {de Laverny}, {De Luise}, {De March}, {De Ridder}, {de Souza}, {de Torres}, {del Peloso}, {del Pozo}, {Delbo}, {Delgado}, {Delisle}, {Demouchy}, {Dharmawardena}, {Di Matteo}, {Diakite}, {Diener}, {Distefano}, {Dolding}, {Edvardsson}, {Enke}, {Fabre}, {Fabrizio}, {Faigler}, {Fedorets}, {Fernique}, {Fienga}, {Figueras}, {Fournier}, {Fouron}, {Fragkoudi}, {Gai}, {Garcia-Gutierrez}, {Garcia-Reinaldos}, {Garc{\'\i}a-Torres}, {Garofalo}, {Gavel}, {Gavras}, {Gerlach}, {Geyer}, {Giacobbe}, {Gilmore}, {Girona}, {Giuffrida}, {Gomel}, {Gomez}, {Gonz{\'a}lez-N{\'u}{\~n}ez},
  {Gonz{\'a}lez-Santamar{\'\i}a}, {Gonz{\'a}lez-Vidal}, {Granvik}, {Guillout}, {Guiraud}, {Guti{\'e}rrez-S{\'a}nchez}, {Guy}, {Hatzidimitriou}, {Hauser}, {Haywood}, {Helmer}, {Helmi}, {Sarmiento}, {Hidalgo}, {Hilger}, {H{\l}adczuk}, {Hobbs}, {Holland}, {Huckle}, {Jardine}, {Jasniewicz}, {Jean-Antoine Piccolo}, {Jim{\'e}nez-Arranz}, {Jorissen}, {Juaristi Campillo}, {Julbe}, {Karbevska}, {Kervella}, {Khanna}, {Kontizas}, {Kordopatis}, {Korn}, {K{\'o}sp{\'a}l}, {Kostrzewa-Rutkowska}, {Kruszy{\'n}ska}, {Kun}, {Laizeau}, {Lambert}, {Lanza}, {Lasne}, {Le Campion}, {Lebreton}, {Lebzelter}, {Leccia}, {Leclerc}, {Lecoeur-Taibi}, {Liao}, {Licata}, {Lindstr{\o}m}, {Lister}, {Livanou}, {Lobel}, {Lorca}, {Loup}, {Madrero Pardo}, {Magdaleno Romeo}, {Managau}, {Mann}, {Manteiga}, {Marchant}, {Marconi}, {Marcos}, {Marcos Santos}, {Mar{\'\i}n Pina}, {Marinoni}, {Marocco}, {Marshall}, {Martin Polo}, {Mart{\'\i}n-Fleitas}, {Marton}, {Mary}, {Masip}, {Massari}, {Mastrobuono-Battisti}, {Mazeh}, {McMillan}, {Messina}, {Michalik},
  {Millar}, {Mints}, {Molina}, {Molinaro}, {Moln{\'a}r}, {Monari}, {Mongui{\'o}}, {Montegriffo}, {Montero}, {Mor}, {Mora}, {Morbidelli}, {Morel}, {Morris}, {Muraveva}, {Murphy}, {Musella}, {Nagy}, {Noval}, {Oca{\~n}a}, {Ogden}, {Ordenovic}, {Osinde}, {Pagani}, {Pagano}, {Palaversa}, {Palicio}, {Pallas-Quintela}, {Panahi}, {Payne-Wardenaar}, {Pe{\~n}alosa Esteller}, {Penttil{\"a}}, {Pichon}, {Piersimoni}, {Pineau}, {Plachy}, {Plum}, {Poggio}, {Pr{\v{s}}a}, {Pulone}, {Racero}, {Ragaini}, {Rainer}, {Raiteri}, {Rambaux}, {Ramos}, {Ramos-Lerate}, {Re Fiorentin}, {Regibo}, {Richards}, {Rios Diaz}, {Ripepi}, {Riva}, {Rix}, {Rixon}, {Robichon}, {Robin}, {Robin}, {Roelens}, {Rogues}, {Rohrbasser}, {Romero-G{\'o}mez}, {Rowell}, {Royer}, {Ruz Mieres}, {Rybicki}, {Sadowski}, {S{\'a}ez N{\'u}{\~n}ez}, {Sagrist{\`a} Sell{\'e}s}, {Sahlmann}, {Salguero}, {Samaras}, {Sanchez Gimenez}, {Sanna}, {Santove{\~n}a}, {Sarasso}, {Schultheis}, {Sciacca}, {Segol}, {Segovia}, {S{\'e}gransan}, {Semeux}, {Shahaf}, {Siddiqui}, {Siebert},
  {Siltala}, {Silvelo}, {Slezak}, {Slezak}, {Smart}, {Snaith}, {Solano}, {Solitro}, {Souami}, {Souchay}, {Spagna}, {Spina}, {Spoto}, {Steele}, {Steidelm{\"u}ller}, {Stephenson}, {S{\"u}veges}, {Surdej}, {Szabados}, {Szegedi-Elek}, {Taris}, {Taylor}, {Teixeira}, {Tolomei}, {Tonello}, {Torra}, {Torra}, {Torralba Elipe}, {Trabucchi}, {Tsounis}, {Turon}, {Ulla}, {Unger}, {Vaillant}, {van Dillen}, {van Reeven}, {Vanel}, {Vecchiato}, {Viala}, {Vicente}, {Voutsinas}, {Weiler}, {Wevers}, {Wyrzykowski}, {Yoldas}, {Yvard}, {Zhao}, {Zorec}, {Zucker}, \& {Zwitter}}]{Gaia2023}
{Gaia Collaboration}, {Vallenari}, A., {Brown}, A.~G.~A., {et~al.} 2023, \aap, 674, A1, \dodoi{10.1051/0004-6361/202243940}

\bibitem[{{Hoogerwerf} {et~al.}(2001){Hoogerwerf}, {de Bruijne}, \& {de Zeeuw}}]{Hoogerwerf01}
{Hoogerwerf}, R., {de Bruijne}, J.~H.~J., \& {de Zeeuw}, P.~T. 2001, \aap, 365, 49, \dodoi{10.1051/0004-6361:20000014}

\bibitem[{{Ivanova} {et~al.}(2005){Ivanova}, {Belczynski}, {Fregeau}, \& {Rasio}}]{Ivanova2005}
{Ivanova}, N., {Belczynski}, K., {Fregeau}, J.~M., \& {Rasio}, F.~A. 2005, \mnras, 358, 572, \dodoi{10.1111/j.1365-2966.2005.08804.x}

\bibitem[{{Lada} \& {Lada}(2003)}]{LadaLada03}
{Lada}, C.~J., \& {Lada}, E.~A. 2003, \araa, 41, 57, \dodoi{10.1146/annurev.astro.41.011802.094844}

\bibitem[{{Lamb} {et~al.}(2016){Lamb}, {Oey}, {Segura-Cox}, {Graus}, {Kiminki}, {Golden-Marx}, \& {Parker}}]{Lamb16}
{Lamb}, J.~B., {Oey}, M.~S., {Segura-Cox}, D.~M., {et~al.} 2016, \apj, 817, 113, \dodoi{10.3847/0004-637X/817/2/113}

\bibitem[{{Leonard}(1995)}]{Leonard1995}
{Leonard}, P. J.~T. 1995, \mnras, 277, 1080, \dodoi{10.1093/mnras/277.3.1080}

\bibitem[{{Leonard} \& {Duncan}(1988)}]{Leonard88}
{Leonard}, P. J.~T., \& {Duncan}, M.~J. 1988, \aj, 96, 222, \dodoi{10.1086/114804}

\bibitem[{{Massey}(2002)}]{Massey02}
{Massey}, P. 2002, \apjs, 141, 81, \dodoi{10.1086/338286}

\bibitem[{{O'Connor} \& {Ott}(2011)}]{OConnor2011}
{O'Connor}, E., \& {Ott}, C.~D. 2011, \apj, 730, 70, \dodoi{10.1088/0004-637X/730/2/70}

\bibitem[{{Oey} {et~al.}(2004){Oey}, {King}, \& {Parker}}]{Oey04}
{Oey}, M.~S., {King}, N.~L., \& {Parker}, J.~W. 2004, \aj, 127, 1632, \dodoi{10.1086/381926}

\bibitem[{{Oey} {et~al.}(2018){Oey}, {Dorigo Jones}, {Castro}, {Zivick}, {Besla}, {Januszewski}, {Moe}, {Kallivayalil}, \& {Lennon}}]{PaperI}
{Oey}, M.~S., {Dorigo Jones}, J., {Castro}, N., {et~al.} 2018, \apjl, 867, L8, \dodoi{10.3847/2041-8213/aae892}

\bibitem[{{Oh} \& {Kroupa}(2016)}]{OhKroupa}
{Oh}, S., \& {Kroupa}, P. 2016, \aap, 590, A107, \dodoi{10.1051/0004-6361/201628233}

\bibitem[{{Perets} \& {{\v{S}}ubr}(2012)}]{PeretsSubr}
{Perets}, H.~B., \& {{\v{S}}ubr}, L. 2012, \apj, 751, 133, \dodoi{10.1088/0004-637X/751/2/133}

\bibitem[{{Pflamm-Altenburg} \& {Kroupa}(2010)}]{Pflamm-Altenburg10}
{Pflamm-Altenburg}, J., \& {Kroupa}, P. 2010, \mnras, 404, 1564, \dodoi{10.1111/j.1365-2966.2010.16376.x}

\bibitem[{{Renzo} {et~al.}(2019){Renzo}, {Zapartas}, {de Mink}, {G{\"o}tberg}, {Justham}, {Farmer}, {Izzard}, {Toonen}, \& {Sana}}]{Renzo19}
{Renzo}, M., {Zapartas}, E., {de Mink}, S.~E., {et~al.} 2019, \aap, 624, A66, \dodoi{10.1051/0004-6361/201833297}

\bibitem[{{Sukhbold} {et~al.}(2016){Sukhbold}, {Ertl}, {Woosley}, {Brown}, \& {Janka}}]{Sukhbold2016}
{Sukhbold}, T., {Ertl}, T., {Woosley}, S.~E., {Brown}, J.~M., \& {Janka}, H.~T. 2016, \apj, 821, 38, \dodoi{10.3847/0004-637X/821/1/38}

\bibitem[{{Vargas-Salazar} {et~al.}(2020){Vargas-Salazar}, {Oey}, {Barnes}, {Chen}, {Castro}, {Kratter}, \& {Faerber}}]{VargasSalazar2020}
{Vargas-Salazar}, I., {Oey}, M.~S., {Barnes}, J.~R., {et~al.} 2020, \apj, 903, 42, \dodoi{10.3847/1538-4357/abbb95}

\end{thebibliography}
\bibliographystyle{aasjournal}

\end{document}